\documentclass[10pt]{article}
\usepackage{amsmath, amsfonts}
\usepackage{algorithmic}
\usepackage{algorithm}
\usepackage{array}
\usepackage{float}
\usepackage{textcomp}
\usepackage{stfloats}
\usepackage{url}
\usepackage{verbatim}
\usepackage{graphicx}
\usepackage{cite}
\usepackage{pifont}
\usepackage{multirow, multicol}
\usepackage{adjustbox}
\usepackage{hyperref}
\hypersetup{
  colorlinks = true, 
  urlcolor = blue, 
  linkcolor = blue, 
  citecolor = green 
}

\begin{document}
\begin{center}
{\huge Emulators in JINSP}
\end{center}
\centerline{
Lei Zhao,
Miaomiao Zhang,
Lv Zhe}
\centerline{\textit{China Mobile Research Institute}}
\begin{abstract}


\textbf{JINSP}\footnote{\href{jiutian.com}{JiuTian Intelligent Network Simulation Platform}} (Jiutian Intelligence Network Simulation Platform) describes a series of basic emulators and their combinations, such as the simulation of the protocol stack for dynamic users in a real environment, which is composed of user behavior simulation, base station simulation, and terminal simulation. It is applied in specific business scenarios, such as multi-target antenna optimization, compression feedback, and so on. This paper provides detailed descriptions of each emulator and its combination based on this foundation, including the implementation process of the emulator, integration with the platform, experimental results, and other aspects.

\end{abstract}

\section{Introduction}
\label{intro}

The platform \cite{zhao2023design} constructs various types of basic emulators and integrates them based on the requirements of different business scenarios. Users immerse themselves in the simulation platform environment and interact with the platform to invoke simulation capabilities and initiate tasks. For example, users can issue simulation configurations to call relevant simulation capabilities and accomplish simulation functions. Subsequently, users can update simulation configurations based on the simulation results returned by the environment (such as antenna parameters, base station positions, etc.). By iteratively invoking emulator capabilities, users can achieve online model training and inference, allowing them to adapt their understanding capabilities to the environment configurations.

 
Each individual basic emulator is responsible for simulating the fundamental capabilities within the network system and may not be able to meet the simulation requirements of complex business scenarios \cite{guo2022overview}. By combining multiple basic emulators and providing unified services, the platform is able to fulfill the task open and online model training service needs for various business scenarios. Currently, the platform offers three combinations of emulators: 
\begin{itemize}
    \item simulation of the protocol stack for dynamic users in a real environment;
    \item simulation of coverage for dynamic users in a real environment;
    \item link-level channel simulation.
\end{itemize}
As the basic emulators and business scenarios continue to expand, the Intelligent Network Simulation Platform and its users will incorporate more emulators.

In the future, the platform will build more complete and rich application scenarios by improving the orchestration mechanism, expanding interfaces, and introducing performance testing standards.

\section{Basic Emulator}
\label{proposal}
This chapter will introduce the basic emulator and combined emulator processes in detail. Basic emulators mainly include user behavior emulator, large-scale channel emulator and channel emulator. On this basis, simulation functions in more complex scenarios can be achieved by combining different basic emulators.

\subsection{User Behavior Emulator}
User behavior simulation includes trajectory generation and business generation. It can be roughly divided into six modules.

\begin{figure}[htbp]
    \centering
    \includegraphics[width=0.8\linewidth]{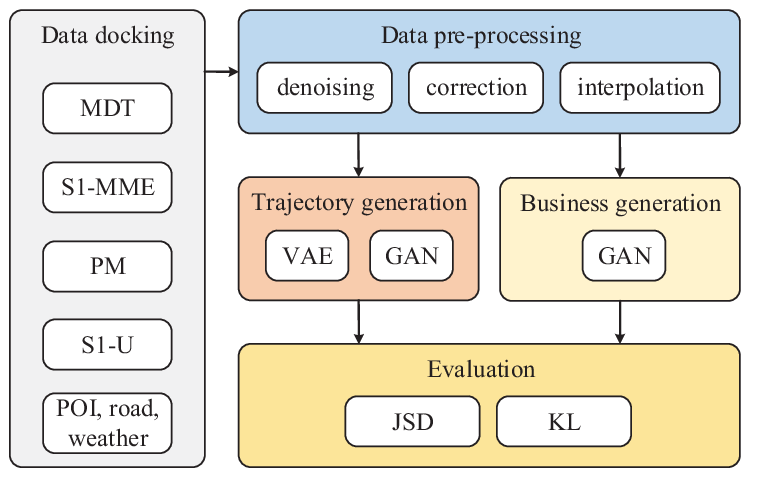}
    \caption{User behavior emulator. MDT: Longitude, latitude and altitude information of some users; S1-MME: multi-user latitude, longitude and altitude information; PM: Community granular user traffic and number of users; S1-U: Full user service type; Other knowledge information: POI, roads, weather, etc.}
    \label{fig:basic-1}
\end{figure}

\begin{itemize}
    \item \textbf{Data docking module} pulls the data available on the existing network, including MDT data, S1-MME data, PM data, S1-U data, etc., and processes it into an algorithm data pre-processing module through extraction and association of relevant fields. Enterable data.
    \item \textbf{Data pre-processing module} performs data analysis based on the user latitude and longitude in the data docking model MDT, the base station latitude and longitude in MME, and user business data in S1-U, including missing, noise, distribution rules, etc., and then perform noise filtering, missing value completion, data correction and other operations, and process it into a format that can be input for offline training of the algorithm and effect evaluation.
    \item \textbf{Offline training module} receives the output data of the data preprocessing module and trains trajectory and business generation models based on VAE \cite{long2023practical, kingma2013auto}, GAN, etc.
    \item \textbf{Online generation module} uses the output model of the offline training module to generate the required user trajectories and business types by configuring relevant parameters online.
    \item \textbf{Data post-processing module} uses map algorithms to interpolate and correct the user trajectories generated by the online generation module to output user trajectories that meet business needs.
    \item \textbf{Evaluation Module} evaluates the effects of trajectory generation and business generation through KL divergence and other indicators to achieve the purpose of generating real and dynamically adjustable data.
\end{itemize} The overall flow chart As shown in Figure \ref{fig:basic-1}.

User trajectories \cite{feng2020learning} have certain periodic regularity in time and space. This feature can be well captured by generation algorithms such as VAE/GAN, which can provide more information based on a batch of original real data. simulated data.

The trajectory generation algorithm based on the VAE scheme includes two modules: Inference and Generation. The Inference module uses the LSTM \cite{yu2019review} coding network to obtain the hidden vector $h$ based on the user identifier $u$, the user's historical location $l$, the historical time $t$, and the user's historical stay time. Based on the latent vector, the parameters of the Gaussian distribution \cite{goodman1963statistical} are obtained through the MLP network, and then the low-dimensional coding vector $z$ is obtained. The Generation module is based on the LSTM decoding network to obtain probability distribution parameters for predicting the user's stay duration and location. Among them, the user's stay duration is described by exponential distribution, and the user's location is described by polynomial distribution. By maximizing ELBO \cite{alemi2018fixing}, the training loss is gradually reduced to achieve the effect of optimizing Inference and Generation.

The overall structure of the GAN scheme follows the core idea of the generative adversarial network and includes two modules, the generator G and the discriminator D. Among them, the generator G contains two sub-networks, namely SeqNet, which uses the Attention mechanism to process time and location sequence information, and RegNet, which processes the user information distribution matrix. Seq can directly generate sequence transcription information based on user locations, and RegNet provides prior knowledge of human mobility patterns by modeling the impact of urban structure on human mobility.

User-level business traffic is also regular and is suitable for generation training using models such as GAN. Its principle is similar to trajectory generation. The difference is that user business data comes from user traffic packets, for example, it is obtained by grabbing and parsing Pcap packets on the network side. Specifically, it includes the following steps:

\begin{itemize}
    \item Obtain subdivided action scene data from the data and extract corresponding fields, such as packet length, arrival time, etc. The clustering algorithm is used to implement statistics for $n$ apps, and obtain the $i$th APP, the number ki of subdivided action categories within $app\_i$, and the center $k\_center$ of the category.
    \item represents the user's feature vector, which is implemented through the knowledge graph to extract user behavior preference information.
    \item Model training stage.
\end{itemize}

\subsection{Large-scale Channel Emulator}
\begin{figure}[htbp]
    \centering
    \includegraphics[width=0.9\linewidth]{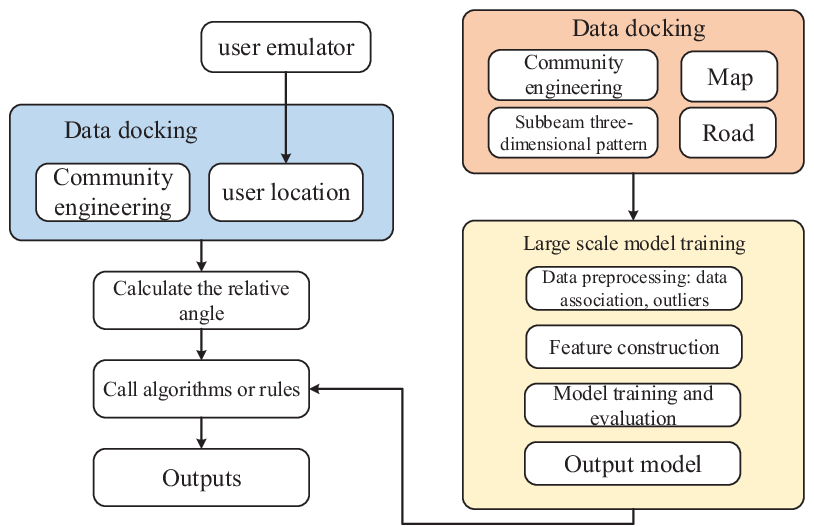}
    \caption{Large-scale channel emulator. The user emulator calculates the relative angle between the user and the sky through the data access module, then calls AI to calculate the large-scale loss, and outputs the large-scale fading between the cell and the user.}
    \label{fig:basic-2}
\end{figure}

According to the antenna pattern of any sub-beam, combined with the location of each cell, antenna height, transmit power, mechanical azimuth angle, downtilt angle, geographical topography, etc., simulate the characteristics of each grid or user's arbitrary cell and any kind of beam. Large-scale channel simulation results, the specific process is shown in the Figure \ref{fig:basic-2}.

\subsection{Channel Emulator}
Based on the real geographical environment, cross-regional and multi-scenario wireless channel modeling is performed based on the regional characteristics of the simulation scene (Indoor, urban micro-cell Umi, urban macro-cell Uma \cite{karttunen2017spatially}, etc.). Channel emulator completes the modeling of wireless channels by calling the basic simulator - large-scale channel emulator, combined with small-scale fading (fast fading) simulation. Among them, the small-scale fading simulation uses a common statistical modeling method to fully simulate the multipath effects caused by direct radiation, reflection, diffraction, transmission, and diffuse scattering during the electromagnetic wave transmission process. Finally, the results of the large-scale fading model are compared with the small-scale fading model. The results are combined to output the simulated channel matrix.

\subsection{Base station/Terminal Emulator}
Base station/terminal simulation provides wireless protocol stack process and function simulation of Radio Resource Control (RRC \cite{zhou2019reinforcement}), Media Access Control Layer (MAC \cite{mwakwata2019narrowband}), and Physical Layer (PHY \cite{o2017introduction}) for system-level simulation. The RRC layer simulation mainly includes cell access, Key processes such as handover reselection and wireless resource management; the MAC layer simulates wireless resource scheduling, multiple input multiple output (MIMO), link adaptation, wireless resource mapping, and uplink power on the base station side and terminal side respectively. Control, Hybrid Automatic Repeat Request (HARQ \cite{ahmed2021hybrid}) and other key processes; while the PHY layer simulation is guided by the above two-layer functional processes and combined with channel simulation to finally calculate the number of access users, cell load, uplink/downlink Multi-dimensional system-level network performance indicators such as transmission rate.

\begin{figure}[htbp]
    \centering
    \includegraphics[width=0.75\linewidth]{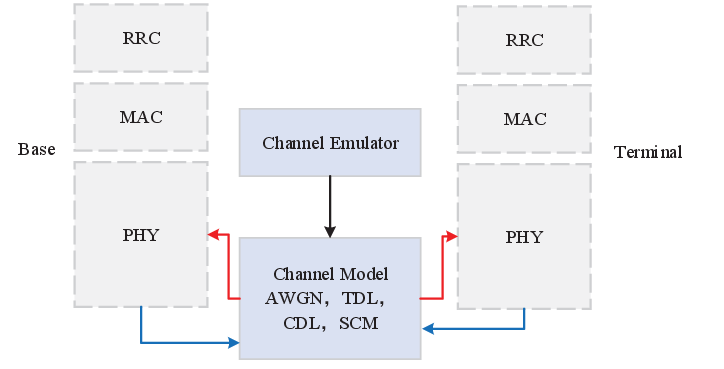}
    \caption{Channel emulator.}
    \label{fig:basic-4}
\end{figure}

On this basis, the base station/terminal simulation module can also be combined with map information, cell parameters, and beam configuration, and use the path loss calculation formula calibrated by the AI algorithm to calculate and output the arbitrary sub-beam antenna pattern for the wireless coverage simulation scenario. Serving cell RSRP/SINR value. The base station/terminal simulation overview diagram is shown in Figure \ref{fig:basic-4}.

\section{Experimental Result}
\label{exper}

Utilizing the Intelligent Network Simulation Platform and interacting with the combination emulator of protocol stack simulation for dynamic users in a real environment, we apply reinforcement learning algorithms to optimize metrics such as RSRP, SINR, traffic, and speed in a real network grid in Jiangxi Province. This process yields a set of more favorable configuration results in the simulation environment, which are then deployed for on-site observation. We conducted measurements on roads and residential buildings in the experimental grid and compared the default configurations of the live network with the optimized configurations generated by the Intelligent Network Simulation Platform. It can be seen from Table \ref{tab:exper_1} \& \ref{tab:exper_2}. that for roads and residential areas, the SS RSRP and SS SINR \cite{afroz2015sinr, park2016analysis} of the overall grid have been improved to varying degrees. Among them, the comprehensive 5G coverage rate of roads increased by 1.79, the downlink rate increased by 106.24 Mbps, and the uplink rate increased by 52.22 Mbps. The uplink rate in residential areas increased by 42.37 Mbps.

\begin{table}[htbp]
    \centering
    \begin{tabular}{cccc}
        \hline
        Metric & Before & After & Delta \\
        \hline
        SS RSRP (dB) & -70.12 & -64.50 & \textbf{5.62} \\
        SS SINR (dB) & 16.46 & 17.41 & 0.94 \\
        5G comprehensive coverage & 98.11\% & 99.90\% & \textbf{1.79\%} \\
        Downlink rate (Mbps) & 788.35 & 894.59 & \textbf{106.24} \\
        Uplink rate (Mbps) & 103.01 & 155.24 & \textbf{52.22} \\
        \hline
    \end{tabular}
    \caption{Comparative effect of road measurement.}
    \label{tab:exper_1}
\end{table}

\begin{table}[htbp]
    \centering
    \begin{tabular}{cccc}
        \hline
        Metric & Before & After & Delta \\
        \hline
        SS RSRP (dB) & -71.12 & -68.89 & \textbf{2.23} \\
        SS SINR (dB) & 12.70 & 14.85 & \textbf{2.15} \\
        5G comprehensive coverage & 100.00\% & 100.00\% & 0.00\% \\
        Downlink rate (Mbps) & 128.15 & 127.66 & -0.49 \\
        Uplink rate (Mbps) & 1056.11 & 1098.48 & \textbf{42.37} \\
        \hline
    \end{tabular}
    \caption{Comparative effect of residential areas.}
    \label{tab:exper_2}
\end{table}

\section{Conclusion}
\label{conclusion}
Through simulation performance optimization, capability combination and interface encapsulation, JINSP provides a good user experience for the external service opening of the simulation network environment and data, and also contributes important infrastructure to the integrated development of communication and artificial intelligence technology.

\bibliographystyle{unsrt}
\bibliography{ref.bib}

\end{document}